\documentclass[aps,prl,twocolumn,groupedaddress,showpacs,amsmath,amssymb]{revtex4}

\pdfoutput=1
\usepackage{graphicx}
\usepackage{dcolumn}
\usepackage{bm}

\newcommand{\mr}[1]{\mathrm{#1}}
\newcommand{\be}{\begin{equation}}
\newcommand{\ee}{\end{equation}}

\newcommand{\kb}{k_{\mr{B}}}
\newcommand{\ls}{L_{\mr{S}}}
\newcommand{\lln}{L_{\mr{N}}}
\newcommand{\lls}{l_{\mr{S}}}
\newcommand{\llo}{L_{\mr{O}}}
\newcommand{\pps}{P_{\mr{S}}}
\newcommand{\rt}{R_{\mr{T}}}
\newcommand{\rn}{R_{\mr{N}}}

\newcommand{\gn}{G_{\mr{N}}}
\newcommand{\gth}{G_{\mr{th}}}
\newcommand{\gthbcs}{G_{\mr{th}}^{\mr{BCS}}}
\newcommand{\gthn}{G_{\mr{th}}^{\mr{N}}}
\newcommand{\ws}{W_{\mr{S}}}
\newcommand{\wn}{W_{\mr{N}}}
\newcommand{\dds}{d_{\mr{S}}}
\newcommand{\ddn}{d_{\mr{N}}}

\newcommand{\dt}{\Delta T}
\newcommand{\asn}{A_{\mr{S}}/A_{\mr{N}}}
\newcommand{\ds}{D_{\mr{S}}}
\newcommand{\lc}{\xi_{0}}
\newcommand{\tho}{\theta_{0}}
\newcommand{\ths}{\theta_{\mr{S}}}

\newcommand{\rr}{\Delta T_2/\Delta T_1}
\newcommand{\lo}{\mathcal{L}_0}
\newcommand{\deltakbt}{y}
\newcommand{\figta}{$\left(\mathrm{a}\right)\;$}
\newcommand{\figtb}{$\left(\mathrm{b}\right)\;$}
\newcommand{\figtc}{$\left(\mathrm{c}\right)\;$}
\newcommand{\figtd}{$\left(\mathrm{d}\right)\;$}
\newcommand{\figa}{$\left(\mathrm{a}\right)$}
\newcommand{\figb}{$\left(\mathrm{b}\right)$}
\newcommand{\figc}{$\left(\mathrm{c}\right)$}
\newcommand{\figd}{$\left(\mathrm{d}\right)$}
\newcommand{\kohm}{\mr{k}\Omega}
\newcommand{\ohm}{\Omega}
\newcommand{\equref}[1]{(\ref{#1})}

\begin{document}

\title{Thermal Conductance of a Proximity Superconductor}

\author{J. T. Peltonen, P. Virtanen, M. Meschke, J. V. Koski, T. T. Heikkil\"a, and J. P. Pekola}
\affiliation{Low Temperature Laboratory, Aalto University, P.O. Box 13500, FI-00076 AALTO, Finland}

\date{\today}

\begin{abstract}
We study heat transport in hybrid normal metal -- superconductor  -- normal metal (NSN) structures. We find the thermal conductance of a short superconducting wire to be strongly enhanced beyond the BCS value due to inverse proximity effect. The measurements agree with a model based on the quasiclassical theory of superconductivity in the diffusive limit. We determine a crossover temperature below which quasiparticle heat conduction dominates over the electron-phonon relaxation.
\end{abstract}


\maketitle

In a bulk superconductor at the lowest temperatures, thermal conductivity is exponentially suppressed compared to the linear temperature dependence expected from the Wiedemann-Franz law \cite{bardeen59}. The residual heat conduction at temperatures $\kb T\ll\Delta$ is only due to quasiparticles at energies above the superconducting energy gap $\Delta$, whereas Andreev reflection completely blocks the sub-gap flow of energy \cite{andreev64}. This is why superconductors are often practically considered as perfect thermal insulators. In hybrid mesoscopic structures with small normal metal islands and short superconducting wires, the picture changes considerably, and heat flow through a superconductor can become essential \cite{courtois08}. When the superconductor (S) is brought into good contact with a normal metal (N) through a transparent metal-to-metal contact, properties of the latter are modified by the widely studied proximity effect \cite{belzig99,gueron96,dubos01,lesueur08}. Close to the interface, also the superconductor is modified by the \emph{inverse} proximity effect: the energy gap is diminished and the sub-gap density of states is non-zero \cite{sillanpaa01}. As a result, the quasiparticle-mediated thermal relaxation through an S wire of length not much larger than the superconducting coherence length $\lc$ is greatly enhanced. Contrary to dying out exponentially at the lowest temperatures, it can dominate over other mechanisms, e.g., electron-phonon relaxation in the N wire.

\begin{figure}[htb]
\begin{center}
\includegraphics[width=\columnwidth]{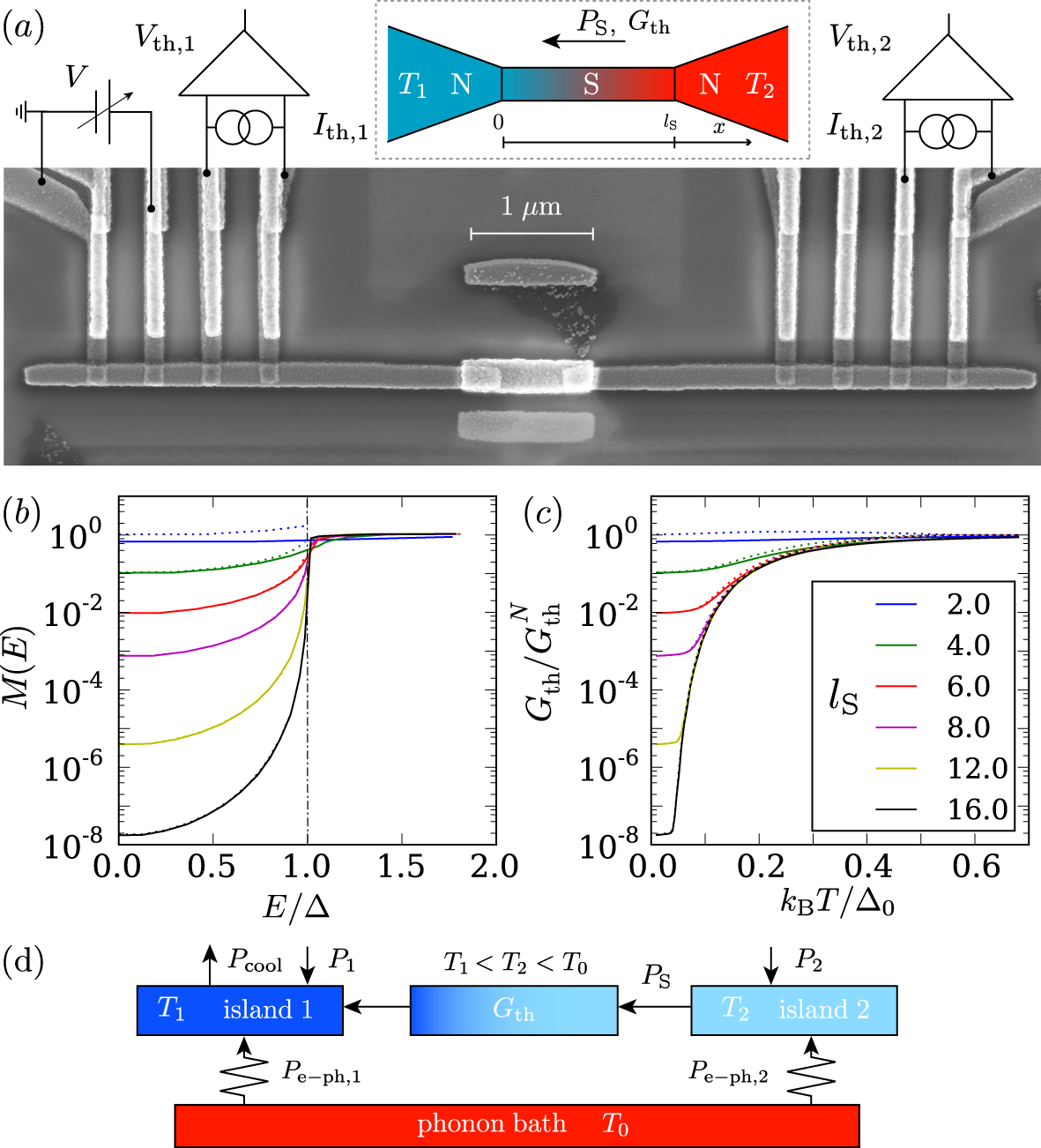}
\caption{(color online). \figta Scanning electron micrograph of a typical sample, together with the configuration for thermal conductance measurements. Two Cu islands are connected via a short superconducting Al wire with transparent NS-interfaces. Four superconducting electrodes (top of the image) are connected to each of the two normal metal islands through tunnel barriers for electronic thermometry and temperature control. Inset: Schematic model of an NSN structure, consisting of a superconducting wire between two normal metal reservoirs at different temperatures.
\figtb Heat transparencies and \figtc thermal conductances from a numerical calculation (solid) and an analytical approximation (dashed) for an NSN structure with an S wire of the indicated length $\lls=\ls/\lc$. \figtd Thermal model for the experimental setup as detailed in the text. Arrows indicate direction of heat flows at temperatures $T_1<T_2<T_0$.}
\label{fig:scheme}
\end{center}
\end{figure}

In this Letter, we report an experimental study of the thermal conductance of diffusive mesoscopic superconducting wires under the influence of inverse proximity effect. This is in contrast to most experiments on Andreev interferometers \cite{chandrasekhar09,eom98,jiang05}, where the thermal conductance depends mainly on the properties of the proximized normal metal, and the focus has been on long range phase coherent effects. Previously, thermal conductance of diffusive NSN structures with short S sections of length $\ls\ll\lc$ was theoretically investigated in \cite{bezuglyi03}. Here, $\lc=\sqrt{\hbar\ds/\Delta_0}$ is the coherence length, $\ds$ is the diffusion constant of the superconductor, and $\Delta_0$ is the energy gap of a bulk superconductor at zero temperature. In this Letter, we focus on S wires with length $\ls\gtrsim \lc$ in the diffusive limit $l\ll\lc,\ls$, where $l$ is the elastic mean free path. Apart from early experiments on large area NSN sandwiches at higher temperatures \cite{deutscher68}, quantitative experiments on thermal conductance of a superconductor affected by the inverse proximity effect are, to our knowledge, missing.

The inset of Fig.~\ref{fig:scheme}\figta shows a sketch of an NSN structure under study. When the N reservoirs are held at different temperatures $T_1$ and $T_2$, heat flow $\pps$ through the superconductor arises. For small temperature differences $\Delta T=T_1-T_2\ll T\equiv(T_1+T_2)/2$, the thermal conductance $\gth$ is defined through $\gth(T)=\pps/\dt$. To calculate $\gth$ within the framework of nonequilibrium superconductivity, pair correlations in the S wire are described in terms of a position and energy dependent complex function $\theta(x,E)$. This pairing angle satisfies the spectral Usadel equation \cite{belzig99,usadel70}
\be
\partial_x^2\theta=-2i\varepsilon\sinh\theta+2i\Delta(x,E)\cosh\theta\label{us1}
\ee
with $\varepsilon=E/\Delta_0$. In Eq.~\equref{us1}, $\Delta(x,E)$ is the self-consistent order parameter in units of $\Delta_0$, and the dimensionless coordinate $x$ is expressed in units of $\lc$. From a solution of the kinetic Usadel equations \cite{belzig99}, it follows that the thermal conductance for a superconducting wire of length $\lls=\ls/\lc$ connected to two N reservoirs via perfectly transparent interfaces is given by
\be
\gth=\frac{\gn}{2\kb T^2e^2}\int_{0}^{\infty}dEE^2M(E)\mr{sech}^2\left(\frac{E}{2\kb T}\right). \label{gth_general}
\ee
Here, $\gn=\rn^{-1}$ denotes the normal state electrical conductance of the S wire, and $M(E)$ is an energy dependent heat transparency defined through $M(E)^{-1}=\lls^{-1}\int_{0}^{\lls}dx\cos^{-2}\left[\mr{Im}\,\theta(x,E)\right]$. In the BCS limit with $\lls\gg 1$, $M(E)=1$ at $E>\Delta$, and it vanishes below the gap. In that case, defining $\deltakbt=\Delta/\kb T$, we recover for $\deltakbt\gtrsim 2$ from Eq.~\equref{gth_general} the result
\be
\gthbcs\simeq2\gn T\left(\kb/e\right)^2\left(\deltakbt^2+2\deltakbt+2\right)e^{-\deltakbt}.\label{gth_bcs}
\ee
On the other hand, in the normal state with $M(E)\equiv 1$, Eq.~\equref{gth_general} reduces to the Wiedemann-Franz value $\gthn=\lo\gn T$ with the Lorenz number $\lo=(\pi^2/3)\kb^2/e^2$.

Neglecting self-consistency of the order parameter and the overlaps of N and S, we can find an analytical approximation for $\gth$ which includes also sub-gap heat transport due to the inverse proximity effect, and describes how $M(E)$ starts to deviate from a step function as $\lls$ decreases. Considering first a semi-infinite NS system with the interface at $x=0$ and the S wire extending along the positive $x$-axis, Eq.~\equref{us1} admits a solution $\theta_{\mr{NS}}(x,E)=\ths-4\mr{artanh}\left[\exp\left(-x\sqrt{2\alpha}\right)\tanh\left(\left(\ths-\tho\right)/4\right)\right]$ with $\alpha=\sqrt{\Delta^2-E^2}/\Delta_0$. Here, $\ths=\mr{artanh}(\Delta/E)$ and $\tho$ are the values of $\theta$ far in the superconductor and close to the interface, respectively. The value of $\tho$ is found by considering the Kupriyanov-Lukichev boundary condition \cite{kupriyanov88} for $\theta$ at the NS-boundary. In the limit of vanishing interface resistance we obtain
\be
\tanh\frac{\ths-\tho}{2}=\frac{\sinh\frac{\ths}{2}}{\cosh\frac{\ths}{2}+ru\left(1-\Delta^2/E^2\right)^{1/4}}.
\label{klboundary2}
\ee
Here, $r=A_{\mr{S}}\sigma_{\mr{S}}^{\mr{N}}/A_{\mr{N}}\sigma_{\mr{N}}^{\mr{N}}$ includes the cross sections and normal state conductivities of the S and N parts. Moreover, the factor $u=\tanh\left[\sqrt{-2i\varepsilon}\lln/\lc\left(2/\ths\right)\sinh\left(\ths/2\right)\right]$ accounts for the finite length $\lln$ of the N part. For a long NSN wire, we obtain the relevant approximate solution as the superposition $\theta_{\mr{NSN}}(x,E)\simeq\theta_{\mr{NS}}(x,E)+\theta_{\mr{NS}}(\lls-x,E)-\ths$. At energies $E<\Delta$ we find
\be
M(E)\simeq 32\left[\mr{Im}\tanh\left(\left(\ths-\tho\right)/4\right)\right]^2be^{-b}\coth\left(b/2\right)\label{me_approx}
\ee
with $b=\lls\sqrt{2\alpha}$, whereas $M(E)\simeq 1$ for $E>\Delta$. This result with $\asn\ll 1$ is compared to non-self-consistent numerical estimates in Fig.~\ref{fig:scheme}\figb, and we see it to be valid for $\lls\gtrsim 4$. The thermal conductance in Fig.~\ref{fig:scheme}\figtc is consequently obtained by using this $M(E)$ in Eq.~\equref{gth_general}, and assuming a BCS temperature dependence for $\Delta$. This is shown below to be in fair agreement with experiments, but especially at higher temperatures and for shorter samples self-consistent numerical calculations become necessary.

In the zero temperature limit the normalized thermal conductance $\gth/\gthn$ saturates to the constant value $M(0)$, and grows as $\sim T^2$ at low temperatures.
For $\asn\ll 1$ and $\lls\gtrsim 4$, we find $M(0)\simeq 32(3\sqrt2-4)\lls\exp(-\sqrt{2}\lls)$. Comparing the sub-gap and above-gap contributions of $\gth$ to each other and to the electron-phonon thermal conductance $G_{\mr{e-ph}}$ of an N wire, one can estimate the significance of the inverse proximity effect to heat transport. For an island of volume $\mathcal{V}$ consisting of a metal with the electron phonon coupling constant $\Sigma$, $G_{\mr{e-ph}}=5\Sigma\mathcal{V}T^4$ \cite{roukes85}. As an example, consider an S wire with $\Delta_0=200\;\mr{\mu eV}$, $\lls=8$ and $\rn=5\;\ohm$, and a copper island with $\Sigma\simeq 2\times 10^9\;\mr{WK}^{-5}\mr{m}^{-3}$ \cite{giazotto06} and $\mathcal{V}=50\;\mr{nm}\times 200\;\mr{nm}\times 1\;\mu\mr{m}$ as the normal metal. Assuming $r=1$, the thermal conductance due to sub-gap quasiparticles becomes equal to the above-gap contribution at a crossover temperature $T_{\mr{cr}}\simeq 170\;\mr{mK}$, and remains dominant over the electron-phonon channel of the N island at lower temperatures.

\begin{table}
\caption{\label{tab:sampletable}Sample parameters, see text for details.}
\begin{ruledtabular}
\begin{tabular}{llcccc}
\multicolumn{2}{c}{Sample} &         I&        II \footnote{Metals deposited in a different evaporator for this structure.}&        III&         IV\\
\hline
$\Delta_0$  &            $\left[\mu\mr{eV}\right]$  &       190&       230&        185&        185\\
$\ls$       &        $\left[\mu\mr{m}\right]$       &       4.2&      1.1 &      0.875&      0.425\\
$\rn$       &        $\left[\ohm\right]$            &       15 &      20  &      5    &          2\\
$\lls$      &                                       &       30 &        8 &        6.5&          4\\
\end{tabular}
\end{ruledtabular}
\end{table}

To probe $\gth$ experimentally, we have fabricated a series of structures similar to the one in Fig.~\ref{fig:scheme}\figa, which displays a typical sample together with the measurement scheme. The structures consist of two normal metal copper (Cu) islands of length $\lln\simeq 2.5-4\;\mu\mr{m}$, width $\wn\simeq 200-250\;\mr{nm}$ and thickness $\ddn\simeq 25-30\;\mr{nm}$, connected by a short superconducting aluminium (Al) wire of width $\ws\simeq 300-400\;\mr{nm}$ and thickness $\dds\simeq 40-50\;\mr{nm}$, with the length $\ls$ varying from sample to sample. The N and S have overlaps of length $\llo\simeq 200-300\;\mr{nm}$. Based on resistivity measurements, we estimate $\ds\simeq 50-75\;\mr{cm}^2/\mr{s}$. Together with the energy gap $\Delta_0\simeq 200\;\mr{\mu eV}$ for Al, this corresponds to $\lc\simeq 100-150\;\mr{nm}$. The samples were fabricated on an oxidized silicon substrate by electron beam lithography and three-angle shadow evaporation of the metals through a suspended resist mask. The structures were measured through filtered signal lines in a \mbox{$^3$He--$^4$He} dilution refrigerator with a base temperature below 50 mK. Here, we present measurements on four samples with the non-overlapped superconductor length $\ls$ varying in the range $400\;\mr{nm}-4\;\mu\mr{m}$. We refer to Table~\ref{tab:sampletable} for sample parameters and dimensions. We estimate the interface resistance of the direct transparent NS-contacts to be less than $1\;\Omega$. Due to strong electron-electron interaction in copper, we assume a well-defined local electronic temperature to exist on each island. Because of the relatively small size of the islands we are able to probe and control this temperature of the electrons.

\begin{figure}[htb]
\begin{center}
\includegraphics[width=\columnwidth]{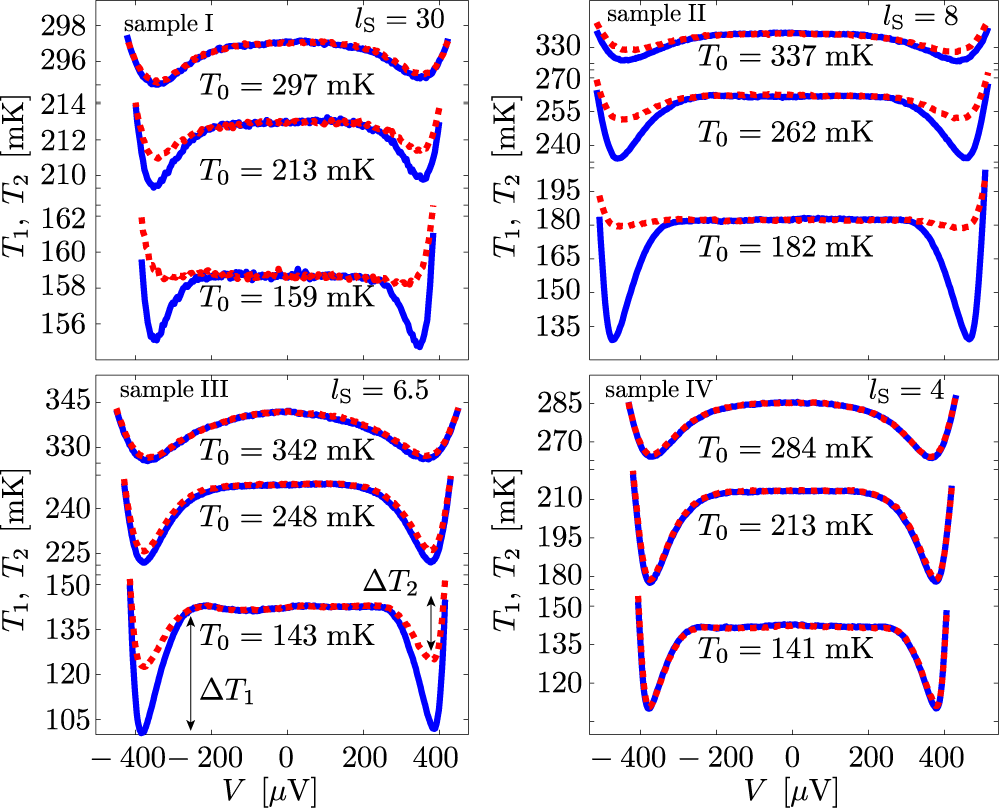}
\caption{(color online). Measured electronic temperatures $T_{1}$ (blue solid lines) and $T_{2}$ (red dotted lines) as a function of the voltage $V$ across the SINIS refrigerator. Each panel shows data acquired at three different bath temperatures $T_0$. Sample I was refrigerated with only a single NIS junction, hence the voltage axis was scaled up by a factor of two.}
\label{fig:coolingcurves}
\end{center}
\end{figure}

Besides the direct NS-contact to the short Al wire, each N island is connected via aluminium oxide tunnel barriers to four superconducting Al leads. These normal metal -- insulator -- superconductor (NIS) tunnel junctions with an area of $(150-200)\times 200\;\mr{nm}^2$ and typical normal state resistance $\rt\simeq 20-100\;\kohm$ facilitate measurement of the electronic temperatures $T_1$ (island 1) and $T_2$ (island 2), and creation of a gradient $\Delta T$ in the following way: As shown in Fig.~\ref{fig:scheme}\figa, on each island  $i$ one pair of NIS junctions is biased by a battery powered floating source at a fixed current $I_{\mr{th},i}\lesssim 0.005\Delta_0/e\rt,\;i=1,2$. Since quasiparticle tunneling in a NIS junction and therefore the current--voltage characteristic is strongly dependent on the normal metal temperature \cite{giazotto06}, the voltages $V_{\mr{th},i}$ act as thermometers once calibrated against the cryostat temperature $T_0$ \cite{nahum93}. To create a temperature difference between the islands, the remaining pair of NIS junctions on island 1 is biased by a DC voltage $e|V|\lesssim 2\Delta$, making this SINIS structure function as an electronic refrigerator \cite{nahum94,leivo96} due to energy-selective quasiparticle tunneling. On the other hand, the low bias current of the thermometer does not significantly affect the thermal balance of the island.

Figure~\ref{fig:coolingcurves} displays the measured electronic temperatures $T_i$ for each sample at three representative bath temperatures $T_0$. For all $T_0$ displayed in Fig.~\ref{fig:coolingcurves}, a drop in the temperature $T_1$ of island 1 is evident close to $e|V|\simeq 2\Delta $, where the cooling power of the SINIS refrigerator reaches its maximum. Similarly, in all cases for $e|V|> 2\Delta $ both islands heat rapidly due to hot quasiparticles entering from the SINIS cooler S electrodes. At smaller $V$, the temperature $T_2$ of the remote island first closely follows $T_1$ at the highest bath temperatures displayed, but at lower $T_0$ a strongly $\ls$--dependent difference develops.
At the observed electronic temperatures, thermal conduction through the substrate is weak and relative changes in the two temperatures as a function of $V$ therefore reflect the thermal conductance $\gth$ of the superconducting wire. To characterize this thermal link between the islands, we choose to study the temperature drops $\Delta T_i=T_i(V)-T_i(V=0)$ at the optimal cooler bias voltage as a function of $T_0$. For consistency we performed several measurements on each sample, permuting the pairs of NIS junctions used for thermometry and refrigeration. The ratio $\rr$ has the advantage of being largely insensitive to the cooling power of the refrigerator junctions, i.e., it is unaffected by their $\rt$ or other characteristics.

To analyze the bath temperature dependence of the relative temperature drop $\rr$, we utilize the thermal model of Fig.~\ref{fig:scheme}\figd. Since the bias voltage $V$ of the SINIS refrigerator is swept at a very low rate compared to the electron--phonon relaxation time, the system reaches a thermal steady state at each $V$, corresponding to the heat balance equations $P_{\mr{cool}}-\pps-P_{\mr{e-ph,1}}-P_1=0$ and $\pps-P_{\mr{e-ph,2}}-P_2=0$ for islands 1 and 2, respectively.
We assume the islands to exchange energy via quasiparticle heat conduction along the superconductor, described by $\gth$ and the heat flow $\pps$. In addition, heat is removed from the cooled island, described by the power $P_{\mr{cool}}$ \cite{leivo96}. At the optimal cooler bias voltage, typical values for the measured samples lie in the range $10-100\;\mr{fW}$. Electrons on each island are thermally coupled to the island phonons at $T_0$ via electron--phonon coupling, modeled by the power flows $P_{\mr{e-ph,i}}=\Sigma\mathcal{V}_{i}(T_{0}^5-T_{i}^5)$ \cite{roukes85}. Here, $\mathcal{V}_{i}$ is the volume of island $i$. Finally, the constant terms $P_i\simeq 1\;\mr{fW}$ account for unavoidable parasitic heating from the electrical environment. We assume a low Kapitza resistance between the Cu island and substrate phonons, thereby neglecting any lattice cooling or heating. This allows us to fix the phonon temperature to $T_0$, i.e., the cryostat bath temperature. We neglect also the photonic heat conduction, because of mismatched impedances, as well as electron-phonon coupling within the superconductor due to the short length of the S wires \cite{timofeev09}.

\begin{figure}[htb]
\begin{center}
\includegraphics[width=\columnwidth]{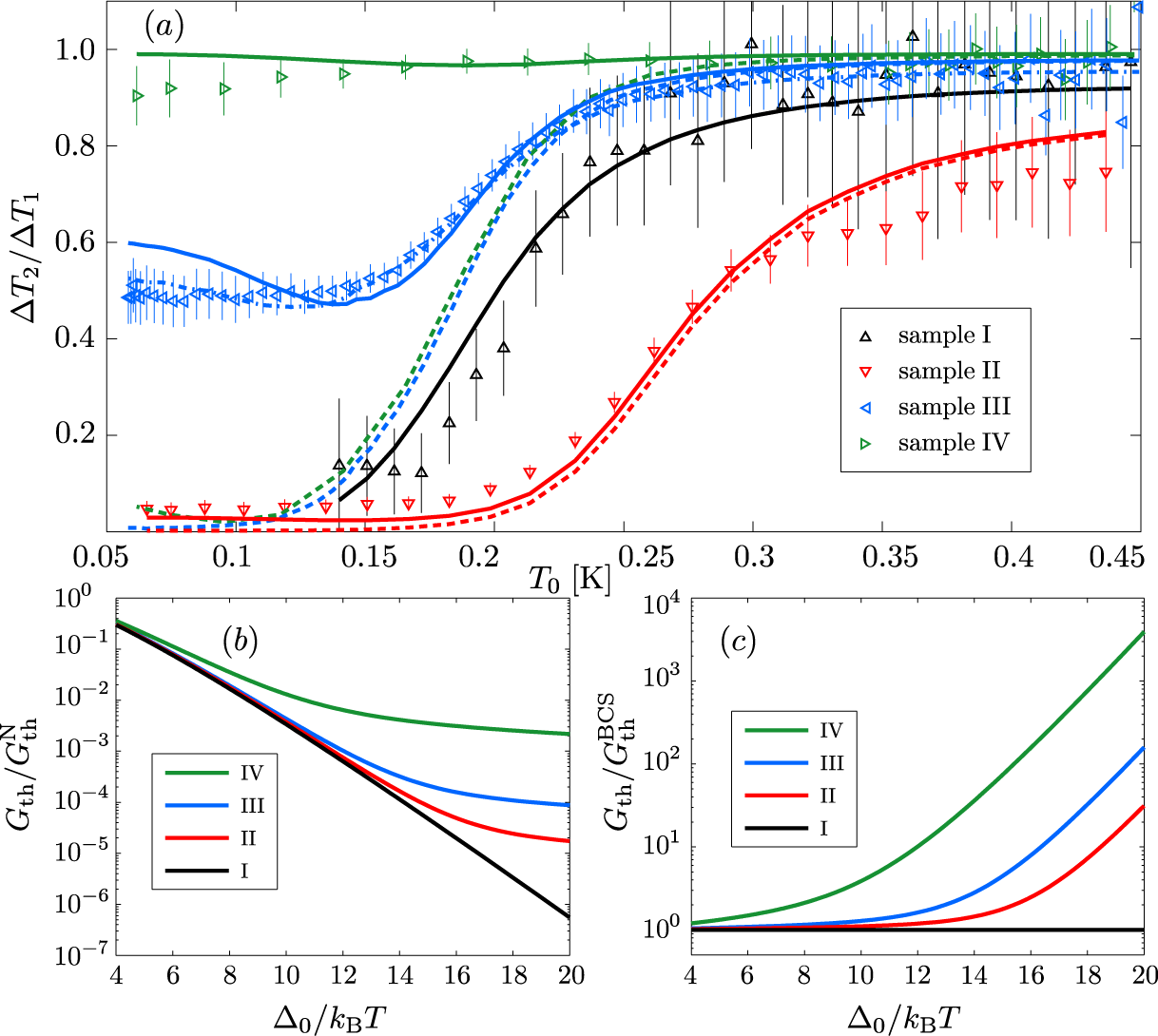}
\caption{(color online). \figta Temperature dependence of the relative temperature drop $\rr$. The symbols show the measured data, whereas the solid, dashed and dash-dotted lines correspond to the thermal model with $\gth$ based on Eq.~\equref{me_approx}, the $\lls\gg 1$ limit of Eq.~\equref{gth_bcs}, and a numerical solution of the Usadel equation, respectively. The error bars are based on the uncertainty in the temperature calibration of the NIS thermometers. \figb,\figtc $T$-dependence of the $\gth$ employed to produce the solid lines in \figa, normalized to $\gthn$ in \figb, and to $\gthbcs$ in \figc.}
\label{fig:deltat}
\end{center}
\end{figure}

Figure~\ref{fig:deltat}\figta displays the measured $T_0$-dependence of $\rr$ for the four samples. Predictions of the thermal model with $\gth$ calculated using Eqs.~\equref{gth_general} and~\equref{me_approx} with $r\simeq 2$ are shown in Fig.~\ref{fig:deltat}~\figta as the solid lines. The dashed lines show the $\lls\gg 1$ limit of Eq.~\equref{gth_bcs}. The lines in Fig.~\ref{fig:deltat}\figtb and \figtc further show $\gth$ relative to its normal state and the BCS limit value, respectively. In all cases, the measured temperatures $T_1$ were used as input for solving the heat balance equation of island 2 to obtain $T_2$, and $\lls$ and $\rn$ were treated as fitting parameters with the values indicated in Table~\ref{tab:sampletable}. Similar results are obtained when the cooling power $P_{\mr{cool}}$ is calculated theoretically, and both heat balance equations are solved. In Fig.~\ref{fig:deltat}\figta the agreement between the model with the analytically approximated $\gth$ and the measurements is reasonable for all the samples. For sample I with largest $\lls$, $\rr$ follows closely the BCS result, similar to sample II with a relatively large $\rn$. Most remarkably, for samples III and IV with the smallest $\lls$, the low temperature behavior of $\rr$ is strongly affected by the inverse proximity. The high values of $\rr$ at the lowest $T_0$ differ drastically from the prediction based on the BCS heat conductance alone (dashed lines). The dash-dotted blue line for sample III is based on $\gth$ obtained from a self-consistent fully numerical solution of Eq.~\equref{us1} in a 1D proximity circuit, including the overlap regions and the series N wires. Comparing to the analytical prediction, the lesser increase in $\rr$ at low temperatures can be partly attributed to an effective increase of $\lls$ due to the proximity effect in the N parts.

To summarize, we have investigated the thermal conductance of short superconducting wires in the presence of inverse proximity effect. We find the conductance to be strongly enhanced relative to that expected for a bulk superconductor. Our study complements earlier work on the thermal conductivity of mesoscopic normal metal wires in close proximity to superconductors. This work helps understanding heat transport in mesoscopic structures, allowing to either utilize or avoid the heat flows through proximized superconductors, e.g., in detector applications of hybrid normal metal -- superconductor structures, or in electronic refrigeration.

\begin{acknowledgments}
We acknowledge financial support from the EU NanoSciERA project "NanoFridge" and the FP7 program "MICROKELVIN". We thank H. Courtois, F. Giazotto and N.B. Kopnin for useful discussions. J.T.P. acknowledges financial support from the Finnish Academy of Science and Letters and T.T.H. from the Academy of Finland.
\end{acknowledgments}

\end{document}